\documentclass[aps,prb,superscriptaddress,showpacs,twocolumn]{revtex4}
\usepackage{amsfonts}
\usepackage{amsmath}
\usepackage{amssymb}
\usepackage{graphicx}

\begin{document}

\title{Enhancement of polarization in a spin-orbit coupling quantum wire
with a constriction}
\author{Jun-Feng Liu}
\affiliation{School of Physics, Peking University, Beijing 100871, China}
\author{Zhi-Cheng Zhong} \affiliation{School of Physics,
Peking University, Beijing 100871, China}
\affiliation{MESA+Institute for Nanotechnology, University of
Twente, 7500 AE Enschede, The Netherlands }
\author{Lei Chen}
\affiliation{School of Physics, Peking University, Beijing 100871,
China}
\author{Dingping Li} \affiliation{School of Physics, Peking
University, Beijing 100871, China}
\author{Chao Zhang}
\affiliation{School of Engineering Physics, University of Wollongong, New South Wales
2552, Australia}
\author{Zhongshui Ma}
\affiliation{School of Physics, Peking University, Beijing 100871, China}

\begin{abstract}
We investigate the enhancement of spin polarization in a quantum
wire in the presence of a constriction and a spin-orbit coupling
segment. It is shown that the spin-filtering effect is
significantly heightened in comparison with the configuration
without the constriction. It is understood in the studies that the
constriction structure plays a critical role in enhancing the spin
filtering by means of confining the incident electrons to occupy
one channel only while the outgoing electrons occupy two channels.
The enhancement of spin-filtering has also been analyzed within
the perturbation theory. Because the spin polarization arises
mainly from the scattering between the constriction and the
segment with spin-orbit coupling, the sub-band mixing induced by
spin-orbit interaction in the scattering process and the
interferences result in higher spin-filtering effect.
\end{abstract}

\maketitle

\medskip

\section{INTRODUCTION}

The generation of a spin-polarized current in low dimensional
semiconductor systems (LDSSs) has been studied extensively both
for fundamental physics and for potential applications in
spintronic devices\cite{wolf}. For these proposes the novel
spintronic materials have been realized. Correspondingly, advanced
electronic
devices, such as spin transistors\cite{datta}, spin waveguides\cite%
{wang}, spin filters\cite{koga}, etc., have been proposed. In
narrow-gap semiconductor nanostructures such as InAs- and
In$_{1-x}$Ga$_{x}$As quantum wells, the inversion asymmetry of the
confining potential due to the presence of the
heterojunction\cite{scha} results in the spin-orbit interaction
(SOI), so that it leads to the spin splitting transport of the
carriers in the absence of any applied magnetic
field\cite{Kakegawa}. It is found that the SOI can be employed to
generate the spin polarization in a T-shaped structure\cite{aa,
masayuki} and a pure
spin current in a Y-shaped junction\cite{tp}. The spin precession\cite%
{francisco,xfwang,hqxu} and spin Hall effect\cite{ha} have been
investigated in quantum wires with SOI. The studies of the
spin-filtering effect in the wire in the presence of SOI show that
the spin-filtering would not occur if the outgoing lead supports
only one open channel\cite{kiselev, xu2}. While for the wire with
multi-channel, a finite transverse nonequilibrium spin
polarization in leads\cite{jiang} is generated. The spatially
averaged polarization density in the transverse direction is found
to be in the third order of the SOI strength\cite{stefano}.

In this work we propose a quantum wire system with a segment of
SOI wire and a constriction to achieve an enhancement of spin
polarization. We shall focus on the spin filtering effect, i.e.,
conductance spin polarization. Based on studies it is a challenge
to enhance the spin polarization significantly in quantum wire
systems in combination structure of a one-channel and a
two-channel sections. It is found that the conductance spin
polarization is very small if the constriction is absent. However,
for the configuration of one channel occupied only in the
constriction and two channels occupied in the segment with SOI,
the spin-polarization reaches up to $95\%$. This is interpreted by
the fact that the higher polarization originates from the
scattering of electrons at the interface between the constriction
and the SOI segment wire. A perturbative analysis show that the
contributions of the lowest two incoming channels to the
spin-filtering always cancel each other partially. Therefore, the
forbiddance of the second incoming channel by the constriction so
that the interference effect between eigenstates of the SOI wire
results in the effect of spin-filtering to be increased.

This paper is organized as follows. In Sec. II  The structure of
SOI quantum wire with a constriction is given. We then present the
scattering-matrix formalism for the multimode spin-dependent
transport. The numerical results of conductances and the
conductance spin polarization are given In Sec. III. To understand
the reason of enhancement of spin polarization, we analyze the
effect of spin-filtering by means of the interference among the
eigenstates of the SOI wire within a perturbative theory. Finally,
a summary is given in the Sec. IV.

\section{Model and Formalism}

\label{sectwo}

\begin{figure}[tbp]
\begin{center}
\includegraphics[bb=9 18 316 186, height=1.853in, width=3.405in]{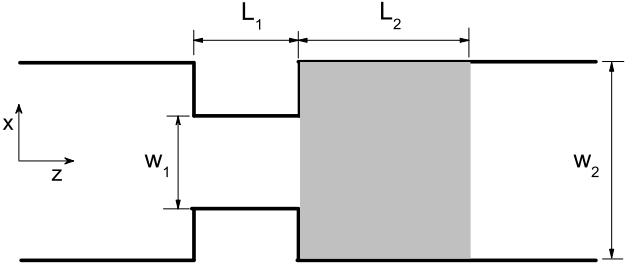}
\end{center}
\caption{Schematic illustration of a quantum wire with a SOI
segment (shaded) and a constriction.} \label{model}
\end{figure}

Considering a two-dimensional electron gas (2DEG) formed in
semiconductor heterostructures, its growth direction is along the
$y$ axis and 2DEG is in the $x$-$z$ plane. In 2DEG a transverse
hard wall potential is applied to form a quasi-one-dimensional
(Q1D) quantum wire of width $W_{2}$. We introduce a constriction
of width $W_{1}$ and length $L_{1}$ in the wire as a schematic
shown in Fig. \ref{model}. With the exception of the segment of
length $L_{2}$ in the presence of the
tunable\cite{nitta,engels,grundler,sato} spin-orbit
coupling\cite{rashba}, no SOI presents in remaining parts of the
wire. The Hamiltonian is given by
\begin{eqnarray}
H &=&\frac{p_{x}^{2}+p_{z}^{2}}{2m^{\ast }}+V(x,z)
+\frac{\lambda \left( z\right) }{\hbar }p_{x}\sigma _{z}\notag \\&-&\frac{1%
}{2\hbar }\left[ \lambda \left( z\right) p_{z}\sigma _{x}+\sigma
_{x}p_{z}\lambda \left( z\right)\right] , \label{hamiltonian}
\end{eqnarray}%
where $m^{\ast }$ is the effective mass of electron, $\lambda
\left( z\right) $ relates to Rashba coupling constant $\alpha $
through $\lambda \left( z\right) =\alpha \left[ \Theta \left(
z-z_{0}\right) -\Theta \left( z-z_{0}-L_{2}\right) \right] $, and
$V(x,z)$ is the confinement potential, i.e.,
\begin{equation}
V(x,z)=\left\{
\begin{array}{cc}
0 & |x|<W_{1}/2\text{ (constriction), or }W_{2}/2\text{ (wire)} \\
\infty  & |x|\geqslant W_{1}/2\text{ (constriction), or }W_{2}/2\text{ (wire)%
}%
\end{array}%
\right.   \label{hardwall}
\end{equation}

The system can be solved by dividing it into several sections,
i.e., the constriction, the segment in the presence of SOI, and
the two semi-infinite wire. Considering the continuous conditions
on the boundaries between sections, the scattering matrix
formalism\cite{andreas} can be built. Because of no SOI in two
semi-infinite wires and the constriction, the solution of
Shr\"{o}dinger equations in these regions are plane waves with
transverse modes due to the confinement in the $x$ direction. For
the segment with finite SOI, we need to consider the solution with
spin-splitting due to the presence of SOI. To do this, we first
introduce the dimensionless units: the coordinate
$\mathbf{x}\rightarrow \mathbf{x}(W_{2}/\pi )$, the energy
$E\rightarrow EE_{0}$ with $E_{0}=\hbar ^{2}\pi ^{2}/(2m^{\ast
}W_{2}^{2})$, the wave vector $\mathbf{k}\rightarrow
\mathbf{k}(\pi /W_{2})$, and the SOI strength $\alpha \rightarrow
\alpha \hbar ^{2}\pi /(2m^{\ast }W_{2})$. In the dimensionless
unit, the Shr\"{o}dinger equation for the segment with a finite
SOI becomes
\begin{equation}
\left[ k_{x}^{2}+k_{z}^{2}+\alpha (k_{x}\sigma _{z}-k_{z}\sigma _{x})\right]
\psi (x,z)=E\psi (x,z).  \label{shrodinger}
\end{equation}%
Its solution can be written in the form of $\psi (x,z)=\varphi
(x)exp ({ik_{z}z})$, where $\varphi (x)$ satisfies the boundary
condition $\varphi (\pi /2)=\varphi (-\pi /2)=0$ due to a
hard-wall potential described by Eq. (\ref{hardwall}) (the
boundaries of wire described by the dimensionless values $\pm \pi
/2$). In general, $\varphi (x)$ can temporarily be written in the
form of $\varphi (x)=\xi_k exp ({ik_{x}x})$, where $k_{x}$ is
determined by the boundary condition and $\xi_k $ is the spinor
which is dependent on wave-vector. For a fixed energy $E$  and a
fixed longitudinal wave vector $k_{z}$, we can obtain four
eigenvalues for $k_{x}$ ($k_{1}$, $k_{2}$, $k_{3}$, $k_{4}$) from
the relation $E=k_{x}^{2}+k_{z}^{2}\pm \alpha
\sqrt{k_{x}^{2}+k_{z}^{2}}$. The
corresponding spinors are given by $\xi _{1}$, $\xi _{2}$, $\xi _{3}$, and $%
\xi _{4}$, respectively. Therefore, $\varphi (x)$ can be expressed
as a superposition of these four eigenfunctions,
\begin{equation}
\varphi (x)=c_{1}\xi _{1}e^{ik_{1}x}+c_{2}\xi _{2}e^{ik_{2}x}+c_{3}\xi
_{3}e^{ik_{3}x}+c_{4}\xi _{4}e^{ik_{4}x},  \label{phix}
\end{equation}%
where $c_{1}$, $c_{2}$, $c_{3}$, and $c_{4}$ are coefficients.
With the help of the boundary conditions at the edges $x=\pm
\pi/2$, we can, in principle, obtain the dispersion relation of
electron, $E=E(k_{z})$, in the wire with the presence of the SOI.
$E(k_{z})$ can only be obtained numerically. Due to the
time-reversal symmetry of the system\cite{xu2}, the dispersion
relation $E(k_{z})$ has the symmetry of $k_{z}$ and $-k_{z}$. For
a fixed value of $k_{z}$ the coefficients $c_{1}$, $c_{2} $,
$c_{3}$, and $c_{4}$ can be obtained by the boundary conditions.
Because of finite width of wire, the transverse modes exist for
the electrons in the wire. In order to reveal possible transport
channels, we use $n$ describes the transverse mode.
Correspondingly, the wave-vector $k_z$ can be written in the form
as $k_z(n,\sigma)$. For example, the analytic relation of energy
can be perturbatively obtained as $E_{n,\sigma}\left( k_{z}\right)
\approx n^{2}+k_{z}^{2}+\sigma\alpha k_{z}-\alpha ^{2}/4$, where
$\sigma=\pm$. Thus a complete set of eigenfunctions for the
quantum wire in the presence of the SOI is found to be
\begin{equation}
\varphi _{n\sigma }(x)e^{ik_{z}(n,\sigma )z}\text{ \ and \ }\tilde{\varphi}%
_{n\sigma }(x)e^{-ik_{z}(n,\sigma )z}.  \label{eigen}
\end{equation}%
It is need to emphasize that the transverse mode $n$ and spin
$\sigma $ are not good quantum numbers. We adopt them as indices
only for the eigenfunctions obtained by numerical calculations.
Generally, the eigenfunctions include the mixing of all the
transverse modes of the wire without SOI. In Section III, the
perturbation theory will indicate that the mixing among the
transverse modes results in the spin-filtering effect. In
addition, the time-reversal symmetry implies that the right-going
and left-going electrons have anti-parallel spins. This leads to
$\varphi _{n\sigma }(x)\neq \tilde{\varphi}_{n\sigma }(x)$.

After obtaining the sets of eigenfunctions for all four sections,
the scattering matrices can be obtained according to the multimode
scattering-matrix procedure\cite{andreas}. In the left (right)
semi-infinite wire without SOI, the wave function can be written
in the form as
\begin{equation}
\psi (x,z)=\sum_{n\sigma }\varphi _{1}^{n\sigma }\left( a_{n\sigma
}^{L(R)}e^{ik_{1}(n,\sigma )z}+b_{n\sigma }^{L(R)}e^{-ik_{1}(n,\sigma
)z}\right) ,
\end{equation}
where $k_{1}(n,+)=k_{1}(n,-)=$ $\sqrt{E-n^{2}}$, and
$\varphi _{1}^{n\sigma }=\sqrt{2/\pi }\sin [n(x+\pi /2)]\left\vert {%
\sigma }\right\rangle _{z}$ with spinors $\left\vert {+}\right\rangle _{z}={\binom{1%
}{0}}$ and $\left\vert {-}\right\rangle _{z}={\binom{0}{1}}$. In
the constriction, the wave function takes the form as
\begin{equation}
\psi (x,z)=\sum_{n\sigma }\varphi _{c}^{n\sigma }\left( a_{n\sigma
}^{c}e^{ik_{c}(n,\sigma )z}+b_{n\sigma }^{c}e^{-ik_{c}(n,\sigma )z}\right) ,
\end{equation}%
where $\varphi _{c}^{n\sigma }=\sqrt{2/\pi r}\sin [n(x+\pi r/2)/r]\left\vert
{\sigma }\right\rangle _{z}$ with $r=W_{1}/W_{2}$, and $%
k_{c}(n,+)=k_{c}(n,-)=$ $\sqrt{E-n^{2}/r^{2}}$. In the segment
with a finite SOI, the wave function is given by
\begin{equation}
\psi (x,z)=\sum_{n\sigma }\left( a_{n\sigma }\varphi _{2}^{n\sigma
}e^{ik_{2}(n,\sigma )z}+b_{n\sigma }\tilde{\varphi}_{2}^{n\sigma
}e^{-ik_{2}(n,\sigma )z}\right) .
\end{equation}

The scattering-matrices can be obtained by considering the
boundary conditions at interfaces, i.e., the continuity of the
wavefunctions and the step change of their derivatives crossing
the interfaces. The step change of the derivatives of
wavefunctions comes from the step change of the strength of SOI
\cite{liu} in the wire. For example, at the interface of the
constriction and the SOI segment, localized at $z=z_{0}$, we have
$ \psi ^{^{\prime }}(x,z_{0}+0_{+})-\psi ^{^{\prime
}}(x,z_{0}-0_{+})=(i\alpha/2)\sigma _{x}\psi (x,z_{0})$. Combining
all the scattering-matrices for all waveguide
sections\cite{andreas}, the total scattering-matrix connects
outgoing amplitudes $\{b_{n_{L}\sigma _{L}}^{L}\}$ and
$\{a_{n_{R}\sigma _{R}}^{R}\}$ to the incoming amplitudes
$\{a_{n_{L}^{^{\prime }}\sigma _{L}^{^{\prime }}}^{L}\}$ and
$\{b_{n_{R}^{^{\prime }}\sigma _{R}^{^{\prime }}}^{R}\}$, i.e.,
\begin{equation}
\binom{b_{n_{L}\sigma _{L}}^{L}}{a_{n_{R}\sigma _{R}}^{R}}=\left(
\begin{array}{cc}
r_{n_{L}\sigma _{L},n_{L}^{^{\prime }}\sigma _{L}^{^{\prime }}} &
t_{n_{L}\sigma _{L},n_{R}^{^{\prime }}\sigma _{R}^{^{\prime }}}^{^{\prime }}
\\
t_{n_{R}\sigma _{R},n_{L}^{^{\prime }}\sigma _{L}^{^{\prime }}} &
r_{n_{R}\sigma _{R},n_{R}^{^{\prime }}\sigma _{R}^{^{\prime }}}^{^{\prime }}%
\end{array}%
\right) \binom{a_{n_{L}^{^{\prime }}\sigma _{L}^{^{\prime }}}^{L}}{%
b_{n_{R}^{^{\prime }}\sigma _{R}^{^{\prime }}}^{R}},
\end{equation}%
where the Einstein's sum rule is adopted and the summation should
be taken over propagating modes in the left (right) semi-infinite
wires\cite{xu2} only. All these propagating modes are normalized
with their velocities. The spin-dependent conductance at zero
temperature is given by summing over the transmission
modes\cite{buttiker},
\begin{equation}
G_{\sigma _{L}\sigma _{R}}=\frac{e^{2}}{h}\sum_{n_{L}n_{R}}|t_{n_{R}\sigma
_{R},n_{L}\sigma _{L}}|^{2}.
\end{equation}%
From this formula, we define the total conductance $G_{\uparrow
}=G_{\uparrow \uparrow }+G_{\uparrow \downarrow }$ for the spin-up
incident electrons , while $G_{\downarrow }=G_{\downarrow \uparrow
}+G_{\downarrow \downarrow }$ for spin-down incident electrons.
The total conductance of non-polarized incident electrons is given
by $G=G_{\uparrow }+G_{\downarrow }$.

The spin polarization after the electrons transporting through the
system (so-called conductance spin polarization) can be expressed
in terms of the scattering-matrix\cite{xu2}:
\begin{eqnarray}
P_{x}+iP_{y} &=&\frac{2e^{2}/h}{G}\sum_{n_{R},n_{L}\sigma
_{L}}t_{n_{R}\uparrow ,n_{L}\sigma _{L}}^{\ast }t_{n_{R}\downarrow
,n_{L}\sigma _{L}},  \notag \\
P_{z} &=&\frac{(G_{\uparrow \uparrow }+G_{\downarrow \uparrow
})-(G_{\downarrow \downarrow }+G_{\uparrow \downarrow })}{G}.
\label{polar}
\end{eqnarray}%
Moreover, because only $x$-component of the conductance spin polarization $%
P_{x}$ is nonzero in present system configuration, we are
concerned with the spin related conductance with respect to the
$x$ direction. The conductances of "spin-up" and "spin-down"
transmitted electrons with respect to the $x$ direction can be
expressed as
\begin{eqnarray}
G_{\sigma _{x}}^{+} &=&\frac{e^{2}}{2h}\sum_{n_{R},n_{L}\sigma
_{L}}\left\vert t_{n_{R}\uparrow ,n_{L}\sigma _{L}}+t_{n_{R}\downarrow
,n_{L}\sigma _{L}}\right\vert ^{2},  \notag \\
G_{\sigma _{x}}^{-} &=&\frac{e^{2}}{2h}\sum_{n_{R},n_{L}\sigma
_{L}}\left\vert t_{n_{R}\uparrow ,n_{L}\sigma _{L}}-t_{n_{R}\downarrow
,n_{L}\sigma _{L}}\right\vert ^{2}.
\end{eqnarray}%
Therefore, the total conductance $G$ satisfies $G=G_{\sigma
_{x}}^{+}+G_{\sigma _{x}}^{-}$ and $P_{x}$ can be defined as%
\begin{equation}
P_{x}=\frac{G_{\sigma _{x}}^{+}-G_{\sigma _{x}}^{-}}{G_{\sigma
_{x}}^{+}+G_{\sigma _{x}}^{-}},
\end{equation}%
which is consistent with Eq. (\ref{polar}). In the next section,
we will perform the numerical calculation and the results show a
significant enhancement of the $x$ component of conductance spin
polarization.

\section{Results and Discussion}

\label{secthree}

In this section we present the numerical calculations of the
conductance and its spin-polarized components. The spin-filtering
will be analyzed in the perturbation theory. In the numerical
calculation, the following parameters are used: $W_{2}=100$ nm,
$m^{\ast }=0.036m_{e}$, $E_{0}\approx 1.04$meV, and $\alpha
_{0}=\hbar ^{2}\pi /(2m^{\ast }W_{2})=3.32\times 10^{-11}$ eVm.

\begin{figure}[tbp]
\begin{center}
\includegraphics[bb=16 14 290 220, height=2.56in,
width=3.405in]{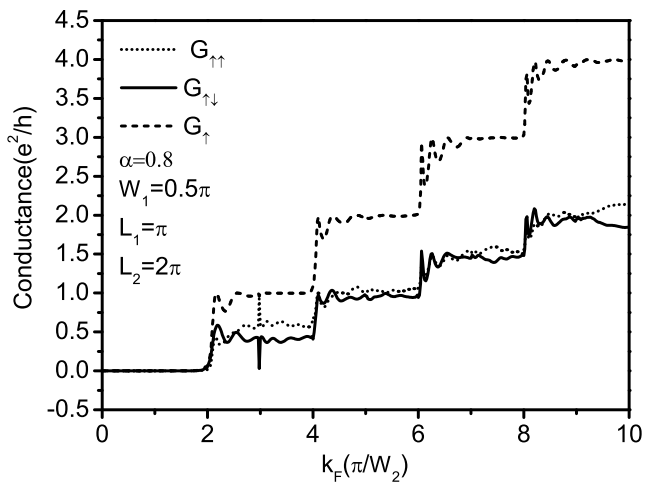}
\end{center}
\caption{Spin resolved conductances $G_{\uparrow\uparrow}$,
$G_{\uparrow\downarrow}$,
and $G_{\uparrow}$ as functions of Fermi wave vector of incident electrons $%
k_{F}=\protect\sqrt{E_{F}}$. The parameters are chosen as (corresponding to $W_2=\protect\pi$): $\protect\alpha%
=0.8$, $W_{1}=0.5\protect\pi$, $L_1=\protect\pi$, and
$L_2=2\protect\pi$.} \label{condu}
\end{figure}

The conductances $G_{\uparrow \uparrow }$, $G_{\uparrow \downarrow }$, and $%
G_{\uparrow }$ as functions of Fermi wave vector of incident electrons $%
k_{F} $ are shown in Fig. \ref{condu}, where $k_{F}=\sqrt{E_{F}}$ with $E_{F}
$ the dimensionless Fermi energy. As it is expected, the total conductance $%
G_{\uparrow }$ is quantized. The appearance of plateaus
corresponds to the energy sweeping over a new channel supplied by
the constriction. The resonant structures near of the edges of
steps come from the
multi-reflection between the two ends of the constriction\cite%
{adstone,george}. In comparison to $G_{\uparrow }$, its
spin-polarized components $G_{\uparrow \uparrow }$ and
$G_{\uparrow \downarrow }$ exhibit more resonant structures
because the Rashba SOI induces the spin splitting and results in a
subband intermixing\cite{francisco}. At the energy near to the
bottom of the third subband $k_{F}\approx 3$, the spin
conductances exhibit a sharp peak for $G_{\uparrow \uparrow }$ and
a sharp dip for $G_{\uparrow \downarrow }$. This phenomena is
related to the details of the spin-dependent scattering mechanism
of electron transport through the whole system configuration.

\begin{figure}[tbp]
\begin{center}
\includegraphics[bb=13 13 293 388, height=4.56in,
width=3.405in]{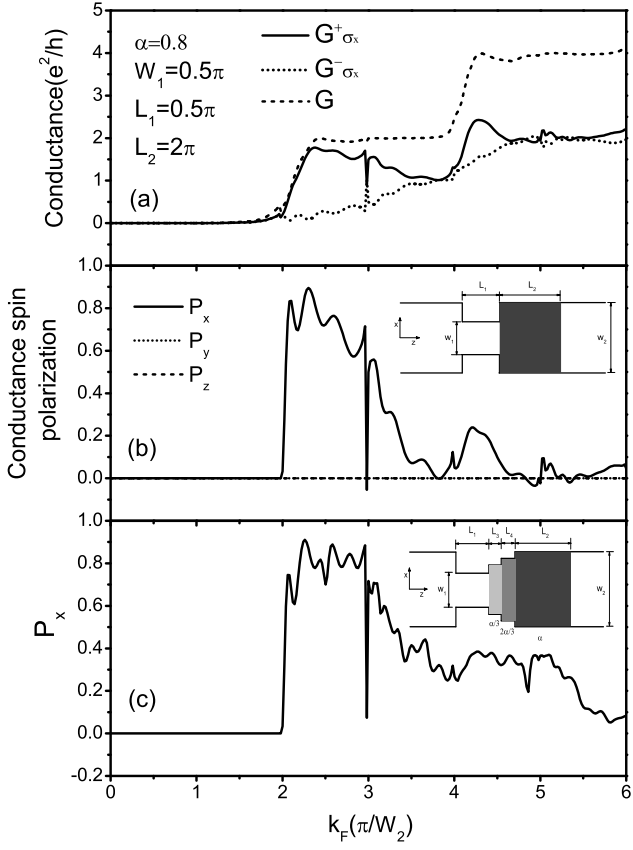}
\end{center}
\caption{(a) Spin resolved conductances $G_{\protect\sigma_x}^+$ and $G_{\protect\sigma%
_x}^-$, and total conductance $G$; (b) The conductance spin polarization $P_x$, $P_y$%
, and $P_z$ as a function of Fermi wave vector of incident
electrons. The parameters are chosen as: $\protect\alpha=0.8$,
$W_{1}=0.5\protect\pi$, $ L_1=0.5\protect\pi$, $L_2=2\protect\pi$.
(c) $P_x$ in the case of sequently increasing widthes and SOI
strengths between the constriction and the SOI segment. The
boundary between the constriction and the SOI segment is smoothed
by adding two short segments with sequently increasing widths
$2\pi/3$, $5\pi/6$ and SOI strengths $0.8/3$, $1.6/3$. The lengths
of these two sections are $\pi$ (as seen in the inset). The other
parameters are the same as those in (b).} \label{polar1}
\end{figure}

In order to show the spin-filtering effect, we calculated the
spin-polarized components of conductance spin polarization. Fig.
\ref{polar1}(a) shows the components $G_{\sigma _{x}}^{+}$,
$G_{\sigma _{x}}^{-}$, and the total conductance $G$. The
conductance spin polarization $P_{x}$, $P_{y}$, and $P_{z}$ as
functions of Fermi wave vector of incident electrons are shown in
Fig. \ref{polar1}(b). It is seen that if the length of the
constriction is chosen to be $L_{1}=0.5\pi =W_{1}$, the plot of
the total conductance $G$ becomes relatively smooth. The reason is
that both the amplitude and the frequency of oscillation depend on
the aspect ratio of the constriction\cite{george}. In Fig. 3(b),
it is found that a large transverse conductance spin polarization
$P_{x}$ can be achieved. The enhancement reaches the strongest in
the situation when the constriction is narrow enough to supply one
channel only and the SOI segment supplies two channels,
\textit{i.e.}, in the range $2<k_{F}<3$. $P_{y}$ and $P_{z}$ are
always vanished because of the symmetry $V(x,z)=V(-x,z)$ and the
time-reversal symmetry\cite{xu2} in the system.

In order to understand how large polarization is generated in this
structure, we analyze the origin of spin-filtering effect in the
perturbation theory. To do this we assume that the SOI is weak and
solve the eigenfunctions of the SOI wire perturbatively. The
Hamiltonian for the SOI segment of wire can be divided into two
parts,
\begin{equation}
H_{0}=k_{x}^{2}+k_{z}^{2}+V(x)-\alpha k_{z}\sigma _{x}
\end{equation}%
and%
\begin{equation}
H_{1}=\alpha k_{x}\sigma _{z}.
\end{equation}%
We then treat $H_{1}$ as a perturbation. The eigenvalues of
$H_{0}$ are found to be $E_{n\pm }^{(0)}=n^{2}+k_{z}^{2}\mp \alpha
k_{z}$ and corresponding eigenfunctions are
\begin{equation}
\varphi _{n\pm }^{(0)}(k_{z},x)=\phi _{n\pm }^{(0)}e^{ik_{z}z},
\end{equation}%
where $\phi _{n\pm
}^{(0)}=\sqrt{{2}/{L\pi }}\sin [n(x+{\pi }/{2%
})]$ with $n$ the transverse subband index , $\left\vert
+\right\rangle _{x}=({1}/{\sqrt{2}})\binom{1}{1}$ and $\left\vert
-\right\rangle _{x}=({1}/{\sqrt{2}})\binom{1}{-1}$ with $\pm $ the
spin index, and $L$ is the length of the
SOI segment. The relevant matrix elements of the perturbation $H_{1}$ are found as
\begin{eqnarray}
\left\langle \varphi _{m\mp }^{(0)}\right\vert H_{1}\left\vert
\varphi _{n\pm }^{(0)}\right\rangle &=&i\left( \alpha /\pi \right)
\left[ 2nm/\left( n^{2}-m^{2}\right) \right]\notag \\
&&[1-(-1)^{n+m}]\delta _{k_{z},k_{z}^{^{\prime }}}
\end{eqnarray}
and other matrix elements are zero. The perturbative eigenfunction
up to the second order in $H_{1}$ is given by
\begin{widetext}
\begin{equation}
\varphi _{n\pm }=\varphi _{n\pm }^{(0)}+\frac{\alpha ^{2}}{\pi ^{2}}%
\sum_{m,l}{}^{^{\prime }}\frac{4nm}{(n^{2}-m^{2})^{4}}\left[
\frac{2nm\varphi
_{n\pm }^{(0)}}{(1\mp \frac{2\alpha k_{z}}{n^{2}-m^{2}})^{2}}+i%
\frac{\pi }{\alpha }\frac{(n^{2}-m^{2})^{2}\varphi _{m\mp }^{(0)}}{%
(1\mp \frac{2\alpha
k_{z}}{n^{2}-m^{2}})}+\frac{4ml(n^{2}-m^{2})^{2}\varphi
_{l\pm }^{(0)}}{(1\mp \frac{2\alpha k_{z}}{n^{2}-m^{2}}%
)(n^{2}-l^{2})(l^{2}-m^{2})}\right] , \label{eigenfunc2}
\end{equation}%
\end{widetext}where the sum extends over the positive integers and $m$ is of
opposite parity with $n$, $l$ is of the same parity as $n$ but
$n\neq l$.

Using these wavefunctions we can calculate the $x$-component of
the spin polarization density. Because the wavefunction of a
right-going electron with definite Fermi energy in the SOI segment
can be written as (assume that only the lowest two channels are
open) $\psi (z)=\sum_{i=1,2}\sum_{\sigma =\pm }a_{i\sigma }\phi
_{i\sigma }e^{ik_{i\sigma }z}$. For simplicity, we ignore the
reflection at the interface between the SOI segment and the right
semi-infinite wire. The $x$-component of the
spin polarization density in the SOI segment is defined as, $%
P_{x}(x,z)=\left\langle \psi (x^{\prime },z^{\prime })\right\vert
\delta (x-x^{\prime })\delta (z-z^{\prime })\sigma _{x}\left\vert
\psi (x^{\prime },z^{\prime })\right\rangle $. It is found
\begin{equation*}
P_{x}(x,z)={Re}\sum_{i,j=1,2}\sum_{\sigma ,\sigma ^{\prime }=\pm
}(a_{i\sigma }^{\ast }a_{j\sigma ^{\prime }}\phi _{i\sigma
}^{\dagger }\sigma _{x}\phi _{j\sigma ^{\prime }}e^{-i(k_{i\sigma
}-k_{j\sigma ^{\prime }})z}).
\end{equation*}

Let us consider an incident electron in the constriction with one
open channel. Because no SOI, the wave function can be written as
$\psi (z)=a_{1+}^{\prime
}\phi_{1+}^{(0)}e^{ik_{1}z}+a_{1-}^{\prime
}\phi_{1-}^{(0)}e^{ik_{1}z}$. Due to $V(x,z)=V(-x,z)$, the
operator $\sigma_{x}R_{x}$ commutes with the Hamiltonian, where
$R_x$ is the reflection transformation with respect to the $x$
axis\cite{xu2}. $\varphi_{1+}^{(0)}$, $\varphi _{1+}$, and
$\varphi_{2-}$ are eigenstates
of $\sigma_{x}R_{x}$ with the eigenvalue $1$. The state $%
\varphi_{1+}^{(0)}$ can only be scattered to $\varphi_{1+}$ and $\varphi_{2-}
$. Similarly, the state $\varphi_{1-}^{(0)}$ can only be scattered to $%
\varphi _{1-}$ and $\varphi _{2+}$ with the eigenvalue $-1$. In
the Landau-B\"uttiker formalism, the conductance is independent of
phases of the incident waves in incoming channels. The phase of
$a_{1+}^{\prime }$ is independent of that of $a_{1-}^{\prime }$.
Therefore, the phases of $a_{1+}$ and $a_{2-}$, which are
transferred from $a_{1+}^{\prime }$, have no relation with those
of $a_{1-}$ and $a_{2+}$, transferred from $a_{1-}^{\prime }$.
Under the phase average, the terms ${Re}\left[ a_{i\sigma }^{\ast
}a_{j\sigma ^{\prime }}\phi _{i\sigma }^{\dagger }\sigma
_{x}\phi _{j\sigma ^{\prime }}e^{-i(k_{i\sigma }-k_{j\sigma ^{\prime }})z}%
\right] $ vanish if $i=j$ and $\sigma \neq \sigma ^{\prime }$, or
$i\neq j$ and $\sigma =\sigma ^{\prime }$.

After taking away those vanished terms under the phase average,
the remained terms in $P_{x}(x,z)$ can be divided into two kinds,
i.e., the contributions from each single state,
\begin{equation}
P_{x}^{s}(x,z)=\sum_{i=1,2}\sum_{\sigma =\pm }\left\vert a_{i\sigma
}\right\vert ^{2}\phi _{i\sigma }^{\dagger }\sigma _{x}\phi _{i\sigma }
\end{equation}%
and the contributions from the interference,
\begin{equation}
P_{x}^{i}(x,z)=\sum_{\substack{ i,j=1,2  \\ i\neq j}}\sum_{\sigma =\pm }{Re}%
\left[ a_{i\sigma }^{\ast }a_{j\overline{\sigma }}\phi _{i\sigma }^{\dagger
}\sigma _{x}\phi _{j\overline{\sigma }}e^{-i(k_{i\sigma }-k_{j\overline{%
\sigma }})z}\right] .  \label{pi}
\end{equation}

\begin{figure}[tbp]
\begin{center}
\includegraphics[bb=13 14 306 275, height=3.033in,
width=3.405in]{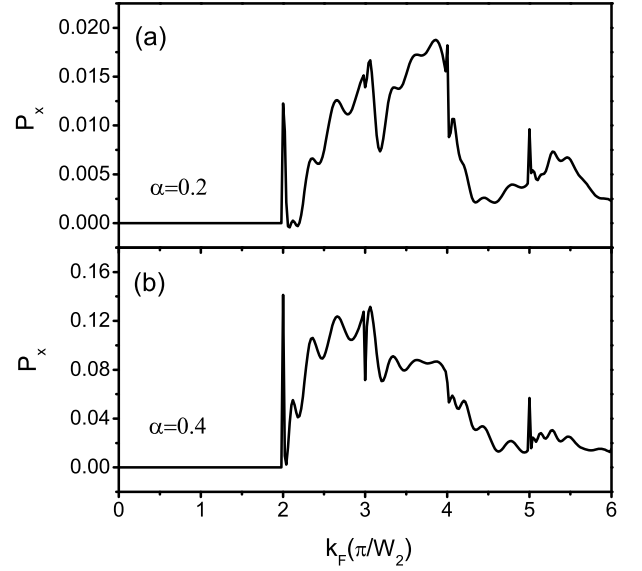}
\end{center}
\caption{Conductance spin polarization for two different values of
SOI strength: $\protect \alpha=0.2$ and $0.4$. The other
parameters are the same as those in Fig. \protect \ref{polar1}.}
\label{polar3}
\end{figure}

Using Eq. (\ref{eigenfunc2}) and integrating over the transverse section of
the wire, we find $\sum_{\sigma =\pm }\int_{-{\pi }/{2}}^{{\pi }/{2}%
}\phi _{i\sigma }^{\dagger }\sigma _{x}\phi _{i\sigma }dx=0+O(\alpha ^{3})$ (%
$i=1$ and $2$). In addition, the dependence of the transmission amplitudes
on $\alpha $ is found as follows: $a_{1+}=[c_{1}+c_{2}\alpha ^{2}+O(\alpha
^{3})]a_{1+}^{\prime }$, $a_{2-}=[c_{3}\alpha +O(\alpha ^{3})]a_{1+}^{\prime
}$, $a_{1-}=[c_{1}+c_{4}\alpha ^{2}+O(\alpha ^{3})]a_{1-}^{\prime }$, and $%
a_{2+}=[c_{5}\alpha +O(\alpha ^{3})]a_{1-}^{\prime }$, where $c_{1}$, $c_{2}$%
, $\cdots $, $c_{5}$ are coefficients determined by the boundary
conditions of the wavefunctions. Therefore, the difference between
$\left\vert
a_{1+}\right\vert ^{2}$ ($\left\vert a_{2+}\right\vert ^{2}$) and $%
\left\vert a_{1-}\right\vert ^{2}$ ($\left\vert a_{2-}\right\vert
^{2}$) is in order of $\alpha ^{2}$. Thus, the contribution from
the single state, $P_{x}^{s}(z)=\int_{-{\pi }/{2}}^{{\pi }/{2}}P_{x}^{s}(x,z)dx$%
, is in order of $\alpha ^{2}$. For those terms contributed from the interference $%
P_{x}^{i}(x,z)$, to the first order, we have $\int_{-{\pi }/{2}}^{{%
\pi }/{2}}\phi _{1+}^{\dagger }\sigma _{x}\phi _{2-}dx=i\alpha
16/\left[ 3\pi (3+2\alpha k_{z})\right] $ and $\int_{-{\pi
}/{2}}^{{\pi }/{2}}\phi _{1-}^{\dagger }\sigma _{x}\phi
_{2+}dx=-i\alpha 16/\left[ 3\pi (3-2\alpha k_{z})\right] $.
Therefore, the contribution from the interference $P_{x}^{i}(z)$
is also in order of $\alpha^2$. As a result, the total
polarization $P_{x}(z)$ is in order of $\alpha ^{2}$. In
comparison to the results in an infinite SOI wire\cite{stefano}
without scattering, where the polarization is in the order of
$\alpha ^{3}$, the $\alpha^2$-dependence of $P_{x}(z)$ in our wire
configuration with a constriction arises from the interface
scatterings. The conductance spin polarizations on different
strength of Rashba SOI are shown in Fig. \ref{polar3}.

\begin{figure}[tbp]
\begin{center}
\includegraphics[bb=13 14 306 221, height=2.305in,
width=3.405in]{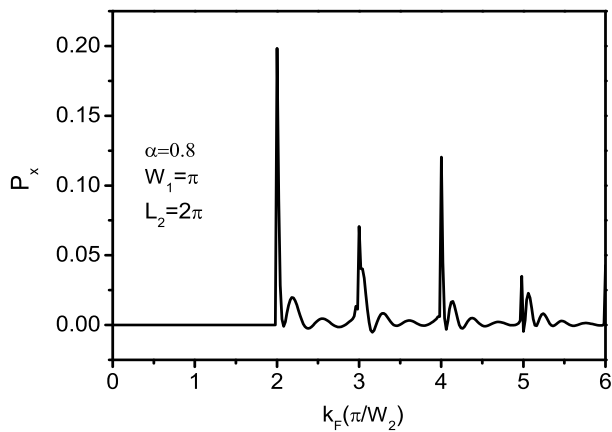}
\end{center}
\caption{Conductance spin polarization for a configuration without
the constriction
($W_1=W_2$). The other parameters are the same as those in Fig. \protect\ref%
{polar1}.}
\label{polar2}
\end{figure}

In order to understand a critical role of the constriction for the
appearance of
a higher spin polarization in this system configuration. We examine $P_{x}^{s}(z)$ and $%
P_{x}^{i}(z)$ separately. For simplicity, we only considered the subband
intermixing among four states $\varphi_{1+}^{(0)}$, $\varphi_{1-}^{(0)}$, $%
\varphi_{2+}^{(0)}$, and $\varphi_{2-}^{(0)}$ induced by the perturbation $%
H_{1}$. Actually, the intermixing  among these four states plays
the main role in achieving a large polarization. For $k_{z}>0$,
the form of the perturbed wavefunctions indicates that the
distortion of $\varphi _{1-}$ and $\varphi _{2+}$ from $\varphi
_{1-}^{(0)}$ and $\varphi _{2+}^{(0)}$ is greater than
that of $\varphi _{1+}$ and $\varphi _{2-}$ from $\varphi _{1+}^{(0)}$ and $%
\varphi _{2-}^{(0)}$. Therefore, the state $\varphi _{1+}^{(0)}$ is more
likely to be scattered into the state $\varphi _{1+}$ than the state $%
\varphi _{1-}^{(0)}$ to be scattered into $\varphi _{1-}$. The probability
of $\varphi _{1+}^{(0)}$ being scattered into the state $\varphi _{2-}$ is
smaller than that of $\varphi _{1-}^{(0)}$ into $\varphi _{2+}$. Namely, $%
\left\vert a_{1+}\right\vert ^{2}$ is larger than $\left\vert
a_{1-}\right\vert ^{2}$ and $\left\vert a_{2+}\right\vert ^{2}$ is
larger than $\left\vert a_{2-}\right\vert ^{2}$. The term
$P_{x}^{s}(z)$ is positive. However, if there is no the
constriction to confine the electron to the lowest channel, the
incident electron can occupy two channels in the
left semi-infinite wire. According to the above analysis, the lowest channel (the states $%
\varphi _{1+}^{(0)}$ and $\varphi _{1-}^{(0)}$) gives a positive
contribution to $P_{x}^{s}(z)$, but the second channel (the states $\varphi
_{2+}^{(0)}$ and $\varphi _{2-}^{(0)}$) will give a negative contribution to
$P_{x}^{s}(z)$. Therefore, the polarization is reduced due to the
contributions of two channels being cancelled partially. Besides, when two
channels are occupied by incident electrons, the interference term $%
P_{x}^{i}(z)$ will be also reduced. For the case without the
constriction, the polarization is plotted in Fig. \ref{polar2}. It
is shown that the polarization is small and nonzero only when the
right lead supplies at least two channels\cite{xu2}. However, the
polarization would be enhanced if there is a constriction which
supplies one channel in the region of constriction and two
channels in the SOI segment(four subbands if spin indexes are
counted). The simplest structure is the width of constriction to
be half of the width of wire $W_{1}=0.5W_{2}$.

\begin{figure}[tbp]
\begin{center}
\includegraphics[bb=16 16 342 222, height=2.152in,
width=3.405in]{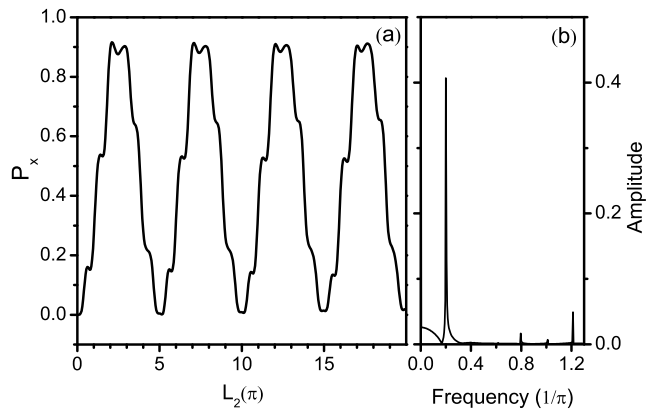}
\end{center}
\caption{(a) The oscillation of conductance spin polarization
along the SOI segment and (b) the corresponding Fourier frequency
spectrum, where the Fermi wave vector is fixed at $k_{F}=2.301$.
The other parameters are the same as those in Fig.
\protect\ref{polar1}.} \label{length}
\end{figure}

\begin{figure}[tbp]
\begin{center}
\includegraphics[bb=13 14 334 242, height=2.418in,
width=3.405in]{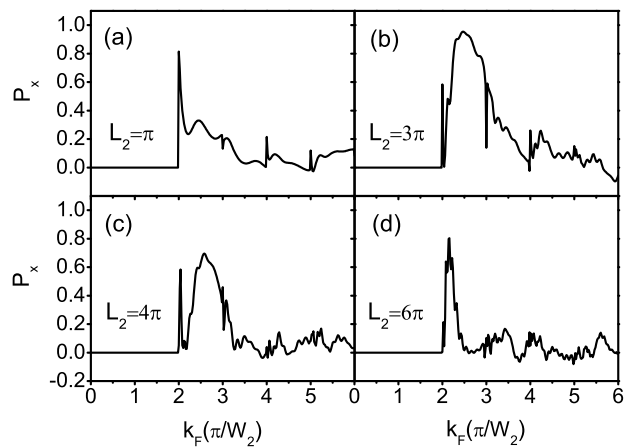}
\end{center}
\caption{Conductance spin polarizations for different lengths of
the SOI
segment: $L_{2}=\protect\pi $, $3\protect\pi $, $4\protect\pi $, and $6\protect%
\pi $. The other parameters are the same as those in Fig. \protect\ref%
{polar1}.} \label{length2}
\end{figure}

Since the above discussion is for $k_{z}>0$, i.e., for right-going
electrons, the main contribution to $P_{x}^{i}(z)$ comes from the
interference between states $\varphi _{1-}$ and $\varphi _{2+}$.
Eq. (\ref{pi}) shows that this interference oscillates with the
longitudinal length in $z$. The period of the oscillation is $2\pi
/(k_{1-}-k_{2+})$. The oscillation of the
polarization $P_x$ along with the length of the SOI segment is plotted in Fig. \ref%
{length}(a). In the numerical calculations the Fermi wave vector is taken $k_{F}=2.301$, so we have $%
k_{1-}\approx 1.890$ and $k_{2+}\approx 1.488$. Therefore, the
period of the oscillation is $4.975\pi $. The structure of the
oscillation with other periods is caused by the interference
between $\varphi _{1+}$ and $\varphi _{2-}$, and the reflection
amplitudes from the right interface of the SOI segment, which can
be seen in the Fourier spectrum Fig. \ref%
{length}(b). For different lengths of the SOI wire, the
conductance spin polarizations as functions of Fermi wave vector
have been shown in Fig. \ref{length2}. The numerical results show
that the polarization is sensitive to the length of the SOI
segment. Especially, when $L_{2}=3\pi$, the large spin
polarization may be kept over a wide range of incident energy.

The above calculations show the effect of the length of SOI
segment on the spin polarization. The width of SOI segment is
influential in the generation of higher spin polarization. Let us
look at the dimensionless form of Hamiltonian first, the
dimensionless coefficient of SOI takes the form as $2m^{\ast
}W_{2}\alpha /(\hbar ^{2}\pi)$ and the energy takes the form as
$2m^{\ast }W_{2}^{2}E/(\hbar ^{2}\pi ^{2})$, where $W_2$ is the
width of the wire. From these we see that the dimensionless values
of the SOI strength and the energy are proportional to $W_2$ and
$W_{2}^{2}$, respectively. In the dimensionless formalism, the
transport of an electron with energy $E$ through a quantum wire of
width $W_{2}$ with the SOI strength $\alpha$ is entirely
equivalent with that of an electron with energy $\gamma ^{-2}E$
through a wire of width $\gamma W_{2}$ with the SOI strength
$\gamma ^{-1}\alpha $, where $\gamma $ is an arbitrary real
number. If the SOI segment is widened ($\gamma >1$) from the width
$W_{2}$ to $\gamma W_{2}$ and and the energy of indicant electrons
is decreased from $E$ to $\gamma ^{-2}E$, but the SOI strength
$\alpha$ is retained unchanged, the only change is the
dimensionless SOI strength is increased from $\alpha $ to $\gamma
\alpha $ in the dimensionless formalism. In another word, when we
widen the wire from $W_{2}$ to $\gamma W_{2}$, keeping the SOI
strength unchanged is equivalent to heighten the dimensionless SOI
strength $\gamma$ times if the energy of indicant electrons is
reduced to $\gamma^{-2} E$ so that the dimensionless energy does
not change. The net effect is to augment the SOI in the region of
segment and leads to enhance the spin polarization. This effect
can be understood qualitatively in another way. As known that the
spaces between the discrete energy levels become small when the
width of the wire increases, so that the contribution from the SOI
is increased. Then, the sub-band mixing becomes stronger. If we
decrease the energy of incident electrons correspondingly so that
scattering from one-channel to two-channel is retained, the higher
polarization is generated.

Although the sharp boundary conditions in the geometry of system
configuration and in the strength of SOI have been considered in
our calculations, both the effect of spin filtering and its
enhancement should be robust and not sensitive to the specific
choice of boundary condition if only a constriction and a SOI
segment are presented. To be compared we smooth the boundary by
adding two short segments with sequently increasing widthes and
SOI strengths between the constriction and the SOI segment. The
numerical results provide the evidences that the amplitude of the
spin polarization is almost unchanged (see Fig. \ref{polar1}(c)).
Only change is the resonant structure. The boundary for the wire
only decides the resonant structure of the conductance spin
polarization, which could be seen in our perturbative analysis,
but not reduce the spin-filtering effect significantly. In
essence, the spin polarization arises from the distortion of the
transverse wavefunction of the SOI segment $\varphi _{n\sigma
}(x)$ from the wavefunction of the wire without SOI $\varphi
_{n\sigma }^{(0)}(x)$ due to the sub-band mixing. The enhancement
of the spin-filtering is less dependent of the boundary. The
alpha-squared scaling for the strength of spin filtering does
remain even in cases with gradual changed boundaries. Differing an
width uniform wire in the presence of a SOI section, in which the
contributions of the lowest two incoming channels to the
polarization of the transmitted current always cancel each other
partially, it is evident that the interference and the scattering
 in the present configuration, due to the configuration of a constriction confining the
incident electrons to occupy one channel only while the outgoing
electrons occupy two channels, increase the spin polarization
significantly.

\section{CONCLUDING REMARKS}

\label{secfive}

In summary, we have investigated numerically the spin-filtering in
a quantum wire with a constriction and a SOI segment. The results
show a higher conductance spin polarization generated in the
transverse direction in comparison to configuration without the
constriction. The spin polarization has been analyzed with the
scattering processes and the perturbation theory. It is found that
the enhancement is mainly due to the presence of a constriction
which confines the incident electrons to occupy only one channel.
The significant spin-filtering occurs when the constriction
supplies one channel and the SOI wire supplies two channels. The
spin-filtering mainly arises from the scattering between the
constriction and the SOI segment. The sub-band mixing effect and
the interference of different spin-states induced by the SOI
segment dominate the spin-filtering effect in the scattering
process. In addition, in contrast to that the polarization is in
the third order of $\alpha $ for a uniform wire, the induced
conductance spin polarization in present system configuration is
in the second order of $\alpha $. The predication of a higher
spin-filtering effect in present configuration is able to be
observed experimentally. The studies have been extended to the
case of gradual change in space between the constriction and the
SOI segment. It is shown that a higher spin-filtering occurs also
if only the constriction supplies one channel and the SOI wire
supplies two channels.

Although the present paper only consider the Rashba SOI, the
significant effect of spin-filtering in the longitudinal direction
instead of in the transverse direction can be retained if the
linear Dresselhaus SOI is utilized instead of the Rashba SOI.
However, the spin-filtering will be reduced if both Rashba and
Dresselhaus SOI are presented. The spin-filtering vanishes if two
kinds of SOI have the same strength. The reason is that the
dispersion relation becomes rigorously parabolic again and the
sub-band mixing disappears.

\begin{acknowledgments}
This work is supported by RFDP-China, NNSF-China No. 10674004, and
NBRP-China No. 2006CB921803.
\end{acknowledgments}

\end{document}